# Two-step phase changes in cubic relaxor ferroelectrics

Eugene V. Colla and M. B. Weissman

Department of Physics

University of Illinois at Urbana-Champaign

1110 West Green Street, Urbana, IL 61801-3080

**Abstract:**

The field-driven conversion between the zero-field-cooled frozen relaxor state and a ferroelectric state of several cubic relaxors is found to occur in at least two distinct steps, after a period of creep, as a function of time. The relaxation of this state back to a relaxor state under warming in zero field also occurs via two or more sharp steps, in contrast to a one-step relaxation of the ferroelectric state formed by field-cooling. An intermediate state can be trapped by interrupting the polarization. Giant pyroelectric noise appears in some of the non-equilibrium regimes. It is suggested that two coupled types of order, one ferroelectric and the other glassy, may be required to account for these data.

**Introduction**

The origins of the glassy 'relaxor' state [1] in cubic perovskite materials such as $PbMg_{1/3}Nb_{2/3}O_3$ (PMN) remain somewhat obscure. [2, 3, 4, 5, 6] Regions with internal ferroelectric order on a scale of about 10 nm, called polar nanodomains, forme at temperatures far above the relaxor regime, e.g. [6]) In the frozen relaxor state these nanodomains acquire slow kinetics but do not form any simple long-range order. In the interesting relaxors, the onset of this frequency-dependent freezing shows much sharper temperature dependence than would be obtained for Arrhenius processes (e.g. [7]), so some thermodynamic cooperativity is involved in the freezing. It is now generally accepted that such relaxors require not just varying strengths of ferroelectric interactions but also some competing terms in the Hamiltonian. Locally antiferroelectric exchange, random vector fields, and random tensor anisotropy all are present to some degree. (e.g. [3]) One key question is whether the nanodomains freeze semi-independently in random fields [8], form fixed constituents of a spinglass-like frozen state[9, 10], represent a typical correlation length within a spherical random-bond picture [3], or whether a second more local type of frozen glassy order arises which couples enough to the polar nanodomains to affect their kinetics[11, 12].

The non-linear response to large electric fields helps to distinguish among different models. When cooled in fields above about 1.6 kV/cm along a [111] direction (these are the easy axes for the ferroelectric polarization), PMN settles into a state with sharper neutron Bragg peaks and reduced diffuse scattering than when zero-field cooled.[13] The extent of polarization of this state does not saturate until around E > 6.5 kV/cm. [13] Zero-field-cooled PMN responds to [111] E-fields above about 1.8 kV/cm by undergoing a first-order phase transition from a globally cubic to a macroscopically polarized phase after a delay time during which the polarization shows roughly logarithmic creep. [14, 15, 16, 17] This time-dependent polarization has been fit fairly well to a mean-field Landau picture together with standard distributed kinetics in the glassy phase.[18] In effect, this picture describes heterogeneous regions, probably polar nanodomains, interacting via a long-range cooperative field (e.g. strain), with a cooperative avalanche occurring when the average value of the ferroelectric order parameter reaches a threshold.

Figure 1 illustrates the standard T-E 'phase diagram' for the standard cubic relaxor PMN. [15]The true equilibrium thermodynamic phase diagram is in general not known, but the boundaries shown represent at least sharp crossovers. In some cases, e.g. the low-field transition or crossover between the paraelectric phase and the glassy relaxor phase, the position of the boundary is nearly independent of history. In other cases, e.g. the transition as a function of E between the glassy relaxor and a state with a ferro-like order, hysteresis is large, evident in the strongly history-dependent position of the boundary in Fig. 1. The Landau analysis suggests that an equilibrium phase boundary lies roughly halfway between these kinetic boundaries. [18]

A study of Brillouin scattering of PMN under applied [111] E-fields shows anomalies at two distinct fields, indicating that the phase diagram of Fig. 1 may fail to capture details of the transition between the glassy and ferroelectric phases.[19] NMR data indicate that the polarized state contains distinct broken symmetry regions along with isotropic glassy regions occupying roughly half the sample.[20] Furthermore, the NMR results seem to show[20] the existence of a first-order transition (but not to a state with *net* broken symmetry) even at E=0 and T≈210 K, i.e. at the boundary between the RXF/FE and simple RXF regimes, not at the boundary indicated by the Landau analysis of the kinetics. These results raise the possibility that the RXF/FE phase transition is more complicated than simple first-order.

Some interesting kinetic effects have been found in the polarized states of PMN. These show Barkhausen noise in the non-linear susceptibility during polarization[14] and in optical birefringence [21] under field reversal, at least in some orientations. Such effects suggest that paying close attention to the kinetics, including noise, of the conversions between phases may be useful.

In this paper we show that in PMN the abrupt, post-creep stage of field-driven conversion between the relaxor and a long-range ferro states actually occurs in two (or sometimes more) distinct, sharp, kinetic stages, as does the thermal conversion at E=0 of that ferro state to the relaxor state. These results support the idea that there are at least two distinct order parameters existing in parallel. As yet the symmetries of the various states are unknown. However, we also find giant non-equilibrium polarization noise which changes strongly at the transitions, facilitating some speculation on the relation of the different transitions to the disordered polarization. Our point here is to present surprising new phenomena rather than to provide a settled explanation.

### Experimental techniques

The bulk of the current data were taken on a single-crystal of PMN, grown by the Czochralski method, obtained from the Institute of Physics at Rostov–on-Don University. A capacitor was constructed with thickness 0.79 mm and area roughly 6 mm$^2$ and with field along a [111] direction. Due to uncertainty in the area, all absolute susceptibilities are uncertain to about 25 %. Brief mention will be made of results from a sample of PMN alloyed with 6% PbTiO$_3$, PMN-6%PT, with field along a [100] direction, obtained from the same source, and of a sample with approximately 12% PT, with [111] orientation.

The main T-E history protocol used was to 1) cool from around T= 350 K in E=0, 2) after about 15 minutes of temperature stabilization at the measurement T, apply sufficient E to drive the system to the FE state and wait for that conversion to occur 3) reset E to zero, and 4) warm at E=0. The state reached after step (2) and maintained after step (3) is denoted ZFC-F2, for reasons which will become clear. In each case, the time (t), the temperature (T), the two components of the dimensionless dielectric constant (ε' and ε"), and the polarization current (I$_P$) were monitored as the system polarized in non-zero E and as it depolarized on warming at E=0. A second protocol was similar except that E was reset to zero immediately after the first peak in I$_P$(t) occurred, giving a state denoted ZFC-F1. A third protocol was to 1) cool at high enough E to induce an FE state, denoted FC, 2) set E = 0, and 3) warm at E = 0. A fourth protocol was simply to cool at E=0, to a state denoted ZF, before warming.

A 1K/min warming rate was used in most cases. Current measurements employed a low-noise current-to voltage amplifier consisting of an AD549LH op-amp operated in inverting mode with a 10 MΩ feedback resistor, in most cases. Although that system produced a small dc offset due to input leakage current of the op-amp, the offset turned out to be highly reproducible at 5.8 nA. Thus the same value has been subtracted from each data set. This baseline uncertainty contributes little to the uncertainty of the net polarizations. Occasional leakage currents give additional baseline uncertainties in polarization data on charging, but not on thermal depolarization. After filtering by a 20 Hz Krohn-Hite 3341 low-pass filter, the signal was digitized by an HP 3478A digital multimeter at a software-determined rate of roughly 0.3 Hz. $\varepsilon'$ and $\varepsilon''$ were measured by applying a 100 Hz voltage, with a low field of about 1.25 V/cm, and measuring the current using the same op-amp current to voltage converter, obviously bypassing the low-pass filter. The applied voltage was used as the reference signal for a standard two-phase lock-in detector. At 100 Hz, phase shifts in the amplification are small, so it was not necessary to take any special precautions to adjust the reference phase for these measurements.

### Results and basic interpretation

In Figure 2, the time dependences of $\varepsilon'$ and $\varepsilon''$ as well as $I_P$ of the device are shown as a function of time delay, t, after applying a field of 3.7 kV/cm to a sample cooled to 175 K in zero field. An initial creep period with a low drift in $\varepsilon$ and a small $I_P$ is found, very similar to that observed in previous work. [14, 16, 17] Then two well-resolved steps in $\varepsilon$ and two peaks in $I_P$ appear. (In some cases, one of these peaks itself split into two sub-peaks.) In contrast, when the sample is cooled in field (to the state FC) it polarizes in a *single* step. Early results [14] on $\varepsilon(t)$ of this sample did not detect the two-step process, apparently because the time resolution was insufficient and no $I_P(t)$ measurements were made. In other prior work on polarization vs. time in PMN (Figure 2 of [17]) it is difficult to see if there were two peaks in $I_P(t)$ because the presentation contained only a few points of the *integral* of $I_P(t)$ during each transition.

For our [111] PMN, the net change of dipole moment (from the integral of $I_P(t)$) is comparable in each of the two steps in Fig. 2, but most of the change in $\varepsilon$ occurs in the second step. In most cases we observed (at various T and E) the change in $\varepsilon$ in the first step was an even smaller fraction of the net change than in the case shown in Fig 2. For example, in a series of runs at T=174 K and E= 3.74 kV/cm, the ratio $(d\epsilon'/dt)/I_p$ in the first peak was 2.5±0.2/nC and in the second peak was 6.7±0.4/nC. Another run, after 15 hours of pre-field aging, gave different values (4.1/nC and 8.7/nC) but a similar contrast for the ratios in the two peaks. Note also in Fig. 2 that the behavior of $\epsilon''$ is dissimilar in the two peaks. We believe the different behaviors of the different measurables in the two peaks is strong evidence that the processes involved in these two steps are dissimilar.

Preliminary work on the [111] PMN-12%PT sample also shows clear multi-step kinetics in both field-induced polarization and thermal melting of the polarization, as seen in Fig. 3. The main qualitative differences between these results and those on PMN are that a sharply defined plateau in $I_P(t)$ precedes the two main peaks, and that there is a smaller trailing peak. Again, only single-step polarization occurred in FC, or in subsequent E=0 heating of FC. We also found that the aging time before applying E affects the delay time before the step-like kinetics. These very time- consuming double-aging experiments will be pursued in future work.

Old data from the [100] PMN-6%PT sample show two sharp features in $d\epsilon'/dt$ during polarization, as shown in Fig. 4. Qualitatively similar results, with two T-dependent times in $d\epsilon'/dt$, were found over several runs in the temperature range of 235 K-240 K. The corresponding $I_P(t)$ data show somewhat more complicated features, with two poorly resolved peaks followed by a plateau. The largest $d\epsilon'/dt$ peak occurs during the relatively small trailing plateau in $I_P(t)$, reminiscent of the data on [111] PMN, for which the change in $\epsilon'$ is also most prominent in the last step.

In each of three cubic relaxor samples we have investigated, the post-creep polarization after ZFC occurs with complicated multi-stage kinetics, as does the subsequent depolarization in the two measured cases. All FC processes are single-step, both on polarization and depolarization. Thus we do not believe that the qualitative result that at least two abrupt stages occur in the field-driven conversion process between the relaxor and ferroelectric states is an artifact of some peculiar sample, although detailed patterns do depend on composition and field direction. For the rest of this paper, we focus on the simple PMN sample, which lacks some of the complications (e.g. mesoscopic domains in ZFC samples and possible doping inhomogeneities) found in PT-doped material.

Although neither step shows instantaneous kinetics, each step occurs within about 10% of the net delay time. That is strongly inconsistent with a collection of domains relaxing independently, which would give at least exponential kinetics or, more realistically in this disordered case, something far broader than exponential. In contrast, the steady $I_P(t)$ and $d\varepsilon/dt$ found preceding the steps would be consistent with some such collection. The obvious interpretation that a first-order phase transition nucleates after sufficient slow domain rearrangements in the glassy phase [16, 17, 18], must be modified to include *two* distinct such transitions.

The two-step conversion occurs for a range of T and E, as shown in Fig. 5. One evident feature was that a large non-equilibrium current noise appeared, both before and even more after the two-step conversion. (Our preliminary results on the PMN-12% PT sample show no such noise.) The E-dependence of the characteristic times, shown in the inset on Fig. 5, can be fit with an exponential dependence, $exp(-E/E_C)$, of the times on E, with characteristic fields $E_C$ of about 180 V/cm.

The T-dependences of the two times, taken at E= 3.17 kV/cm as seen in Fig. 6, fit Arrhenius laws with slightly different activation energies for the first and second step, as described in the caption of Fig. 6. The formal attempt rate for the second step is un-physically high, indicating that the transition state for that process probably has higher entropy than the metastable states.

Since the processes are thermally activated, one can use the E-dependence to calculate a characteristic dipole-moment change $p$ associated with the kinetics: $p = k_B T/E_C$. The resulting value, $p =1.4 \times 10^{-23}$ C-cm, corresponds approximately to the dipole moment of a single nanodomain, confirming the previous speculative picture that the creep phase consists of events on the nanodomain scale. [14, 16, 17, 18]

Figure 7 shows how $\varepsilon$ and $I_P$ change on gradual heating at E=0 after several different treatments. After two-step polarization (to the state ZFC-F2), depolarization also occurs in two distinct steps. The depolarization is significant in each step, but the change of $\varepsilon$ is mainly confined to the second step. In contrast, the relaxation on heating of the FC state (formed in a single step) at E=0 also occurs in *one* step, as seen in Figure 4, at slightly higher T, and with larger net change in $\varepsilon$.

Integrating over the entire thermal depolarization $I_P$, ZFC-F2 shows less polarization loss on melting than does FC. The results were not entirely consistent between runs but clustered around 50% of the FC thermal $I_P$ integral. This result differs somewhat from prior work, in which the field-induced polarization was reported to be equal to the field-cooled polarization. [17] Although we measured the net polarization change on E=0 heating, rather than on isothermal poling, [17], the initial polarization relaxation on setting E=0 was under 10% of the total polarization, not enough to make up the discrepancy. These results then indicate that there can be important sample-dependences.

Figures 5, 7, and 8 illustrate the enormous non-equilibrium current noise accompanying conversions in either direction between the field-induced states and the relaxor state. This noise provides a signature which allows more sensitive discrimination between different metastable states than provided by the average current and susceptibility alone. The anomalously large current noise found on warming the ZFC-F2 state decreases just before or during the current spikes, depending on the run, although a noise tail persists after the spikes.

By turning off the field after the first step but before the second step of the field-induced ordering, we prepared a zero-field state (denoted ZFC-F1) which should closely resemble the intermediate state in the conversion. ZFC-F1 shows even more $I_P$ noise upon warming than does ZFC-F2, as seen in Fig. 8. On all three runs attempted, very similar results appeared, despite a variety of aging times at E=0 before warming. The $I_P$ spike on melting is smaller than either spike for ZFC-F2. The noise reduced substantially at the melting T. However, even above this nominal transition temperature, the $I_P$ noise remains far above the equilibrium value, and above the level shown by ZFC-F2 in the same regime.

One surprising feature of the $I_P$ noise before the melting transition for both ZFC-F2 and ZFC-F1 is that it shows frequent excursions to values with sign opposite to the mean $I_P$. Thus it cannot be modeled as a simple collection of randomly timed de-polarization steps. Some large re-polarization steps must also be present.

The variety of qualitatively distinct states obtainable at E=0 in a narrow range of temperature near 206 K, depending on E-T history, is summarized in this table. ZF is the ordinary relaxor state, obtained by zero field cooling. The peak-to-peak pyro noise numbers are functions of the measurement bandwidth, and are intended only for comparisons with each other. As evident in the figures, they are only reproducible to around a factor of two. The ΔP's given here do not include the pre-or post-peak contributions.

| State | Melting steps at: | ΔP per step ($\mu C/c\,m^2$) | Pyro noise (nA) |
|---|---|---|---|
| ZF | NA | 0 | small |
| FC | ~210 K | ~20 | small |
| ZFC-F2 | ~207.8K ~208.4K | ~2 ~5 | ~1 |
| ZFC-F1 | ~209 K | up to ~2 | ~2 |

## Discussion

Before proceeding with further interpretation, we need to check whether some simple model in which there happen to be two distinct regions in the sample could account for the two-step data. For example, it is known that rather thick surface layers can form different phases than the bulk (e.g. [22, 23]). Accidental inhomogeneities can also occur. Several pieces of evidence indicate that such macroscopic inhomogeneities are not the explanation. First, the melting of FC shows only a single stage, which would not make sense if there were two nearly independent regions. Second, the melting of ZFC-F1 does not closely resemble *either stage* of the melting of ZFC-2, being broader than either sharp step. Third, ZFC-F1 shows more pyrocurrent noise than either of the states it is intermediate between. Fourth, the first current step after the field or on thermal depoling are accompanied with relatively smaller changes in ε' than are the second, which makes little sense if the two steps represent roughly similar processes in different regions. Fifth, only the first step is accompanied by a significant peak in ε". We conclude that whether or not the two sharp stages are primarily associated with different microscopic regions, these stages involve distinct but closely coupled processes, not a sum of two similar but macroscopically disjoint processes.

The two-step conversion processes do not fit a picture with a single well-defined intermediate state for both ordering and disordering, since the intermediate state in melting has ε very close to that of the ferro state, but the intermediate in ordering has ε is close to the relaxor state. Rather, there seem to be two processes, one of which (the one that changes ε more) has larger kinetic barriers, and hence happens second in either the formation or destruction of this ferro state. This second process also appears to have negative activation entropy, perhaps indicative of some partial melting in the transition state.

Given that there are at least two distinct sharp stages of the ferro-relaxor conversion, and at least three distinct types of field-induced states, it remains to examine the possible connections to proposed pictures of the relaxor state. Neutron scattering studies of $(PbZn_{1/3}Nb_{2/3}O_3)_{0.92}(PbTiO_3)_{0.08}$ have shown that for applied fields of 2kV/cm to 10 kV/cm along [001] in the tetragonal phase, the nanodomains with polarization axes parallel to the field appear to align coherently, while displacements orthogonal to the field remain random.[24] The explanation suggested was based on the different magnitudes of field-dipole dot products for different nanodomain principal axes[24], i.e. on a picture involving disorder at a single length-scale, the nanodomain size. Such different types of nanodomain reorientations might also be involved in the two-step polarization kinetics. Such an explanation is not especially obvious, however, since for the [111] PMN sample which usually shows two distinct peaks, there are three categories of <111> domains to realign, with different numbers of negative indices. We suspect that the origin of at least one of the steps may not be closely connected to the large-scale symmetries.

It is also difficult to see what the two abrupt stages of ordering would be for simple models of spinglasses with nanodomain constituents, or for the spherical random bond picture. [3] The principal difficulty of the spherical random bond picture is that it seems to require two different values for a single coupling strength to model the onset of frozen order and the divergence of the non-linear susceptibility[3], already suggesting that a modified two-order-parameter version might be needed.

It was suggested many years ago that the relaxor state was related to some sort of reentrant spinglass. [25] We have proposed that the relative insensitivity of the spinglass-like aging effects to electric field changes as well as the temperature dependence of Barkhausen noise indicate that the frozen glassy order occurs among units with dipole moments much smaller than nanodomains.[11, 12] The unit-cell-scale displacements orthogonal to the ferro moments [2] provide such degrees of freedom, analogous to those of reentrant xy spinglasses.[26, 27] Their glassy ordering could then coexist with ferro order, affecting its kinetics. There is clear evidence that glassy regions exist between nano-domains[16, 20], even in field-aligned material, although it is less clear to what extent frozen glassy order spatially overlaps with the ferro regions. One stage of the kinetics, whether from relaxor to ferro or from ferro to relaxor, could then involve changes in this small-scale order, not mere realignment of nanodomains.

More generally, the interplay of these two different scales of disorder is more likely to be able to provide an explanation for the rich collection of distinct metastable states than is a picture with a single scale. The key ingredients would be that the glassy degrees of freedom have many metastable states and that there are major terms in the Hamiltonian coupling these glassy modes randomly to the ferroelectric orientations. Thus different states of domain orientations would be in equilibrium with different versions of the glassy order. Since the kinetics for these two types of coupled degrees of freedom would be different, two-step kinetics could arise. A key non-trivial aspect of this picture would be that the glassy states themselves could have large-scale cooperative transitions, not just the ferroelectric domain states.

Among the possibilities for extra glass-related kinetic stages are: 1) rearrangement to a new glassy state closer to equilibrium with the domains, 2) collapse of orthogonal components within ferro domains after the overall symmetry is broken 3) growth of ferro domains into previously glassy regions after overall symmetry is broken.

For illustration only, we present some highly speculative interpretations of the kinetics. It is not very likely that all these speculations will turn out to be correct, but we hope they provoke some more serious theoretical work on coupled glass-domain models.

The direct effect of the applied field should be strongest on the most nearly ferroelectric degrees of freedom. We expect that the first sharp transition after the field is applied would involve cooperative large-scale realignment of nanodomains, along the lines previously suggested. [17, 18] The peak in $\varepsilon$" during this step would be a typical symptom of having a lot of domain walls. The intermediate stage (ZFC-F1) would then have ferro domains, but coupled somewhat randomly to persistent glassy degrees of freedom previously frozen in the presence of unaligned domains. In the presence of the new effective Hamiltonian for the small-scale glassy degrees of freedom, a cooperative transition between different glassy states could then occur, accompanied by further macro-realignment, to ZFC-F2. The idea that the nanodomains are out of equilibrium with the glassy modes in ZFC-F1, and to a lesser extent in ZFC-F2, is consistent with the anomalous pre-melting $I_P$ noise found.

The complicated melting of the ZFC-F2 state may also fit a similar picture. The first large $I_P$ peak may involve realignment of nearly independent nanodomains within a nearly fixed glassy matrix. The second peak would involve the relaxation toward equilibrium of the coupled glass-nanodomain system.

Some central questions remain for future work. The first is whether the key ingredient of the relaxor glassy state is the disorder of the nanodomain polarizations or (as we believe is more compatible with the data) of other, smaller-scale, components of the polarization. The second is whether the known disordered components of the polarization exist mainly outside the polar nanodomains (as suggested by NMR [20]) or also coexist with the ferroelectric order within the nanodomains, in analogy to the reentrant xy spinglasses[26, 27], as suggested by neutron scattering pair-density functions[2]. Combining scattering studies with the two-step kinetics should allow more definitive answers to these questions. Despite their poor time resolution, NMR and neutron scattering should be able to reveal the structure of ZFC-F1, the distinct metastable intermediate state which we have shown can be prepared if the polarization is interrupted after one step. It may also turn out to be possible to trap the intermediate state which occurs on warming. The nature of the randomly changing order in noisy regimes may be revealed by x-ray speckle interferometry. It should also be possible to develop model-dependent predictions for the giant polarization noise.

**Acknowledgements:**

This work was funded by NSF DMR 02-40644 and used facilities of the Center for Microanalysis of Materials, University of Illinois, which is partially supported by the U.S. Department of Energy under grant DEFG02-91-ER4543. We thank D. Viehland for providing facilities and assistance for the initial work on the PMN-6%PT sample and S. Vakhrushev for stimulating conversations.

**Figures**

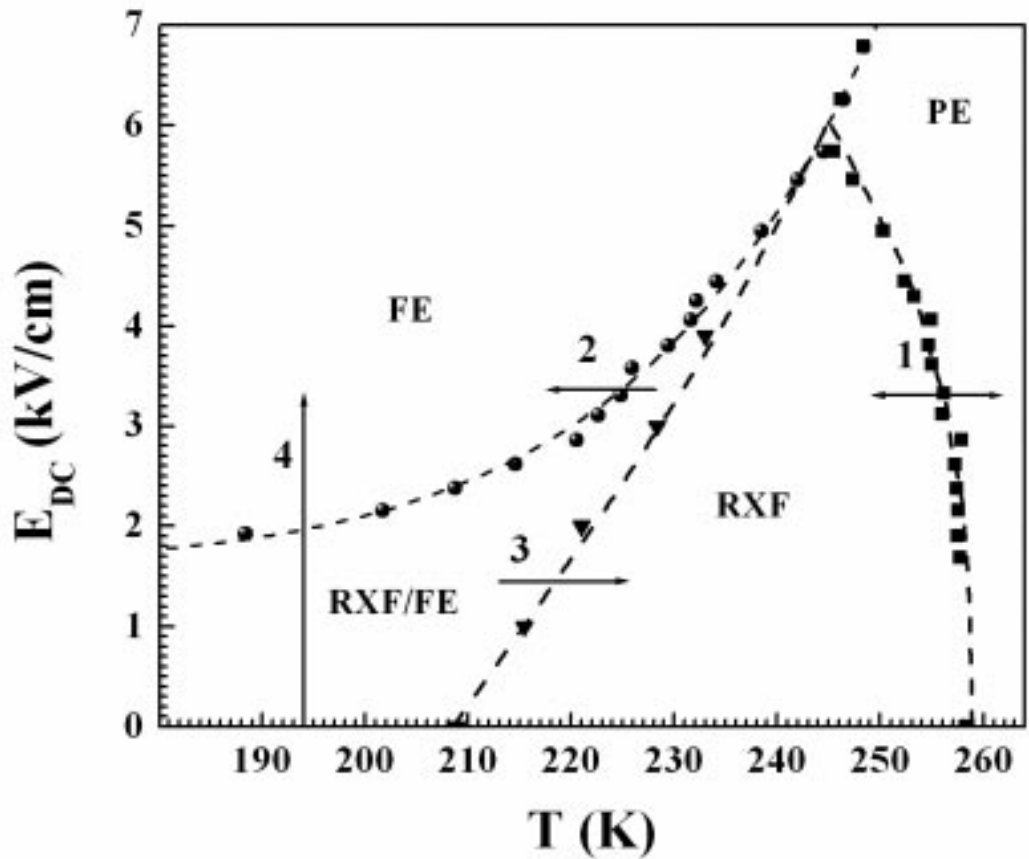

Figure 1. The standard empirical history-dependent phase diagram [15, 16, 17] is shown for PMN. Sharp crossovers are shown with arrows to indicate the conditions (warming, cooling, increasing E) under which they were found. 'FE' represents a ferroelectric state, 'PE' a paraelectric state with nanodomains, and 'RXF' a glassy relaxor state.

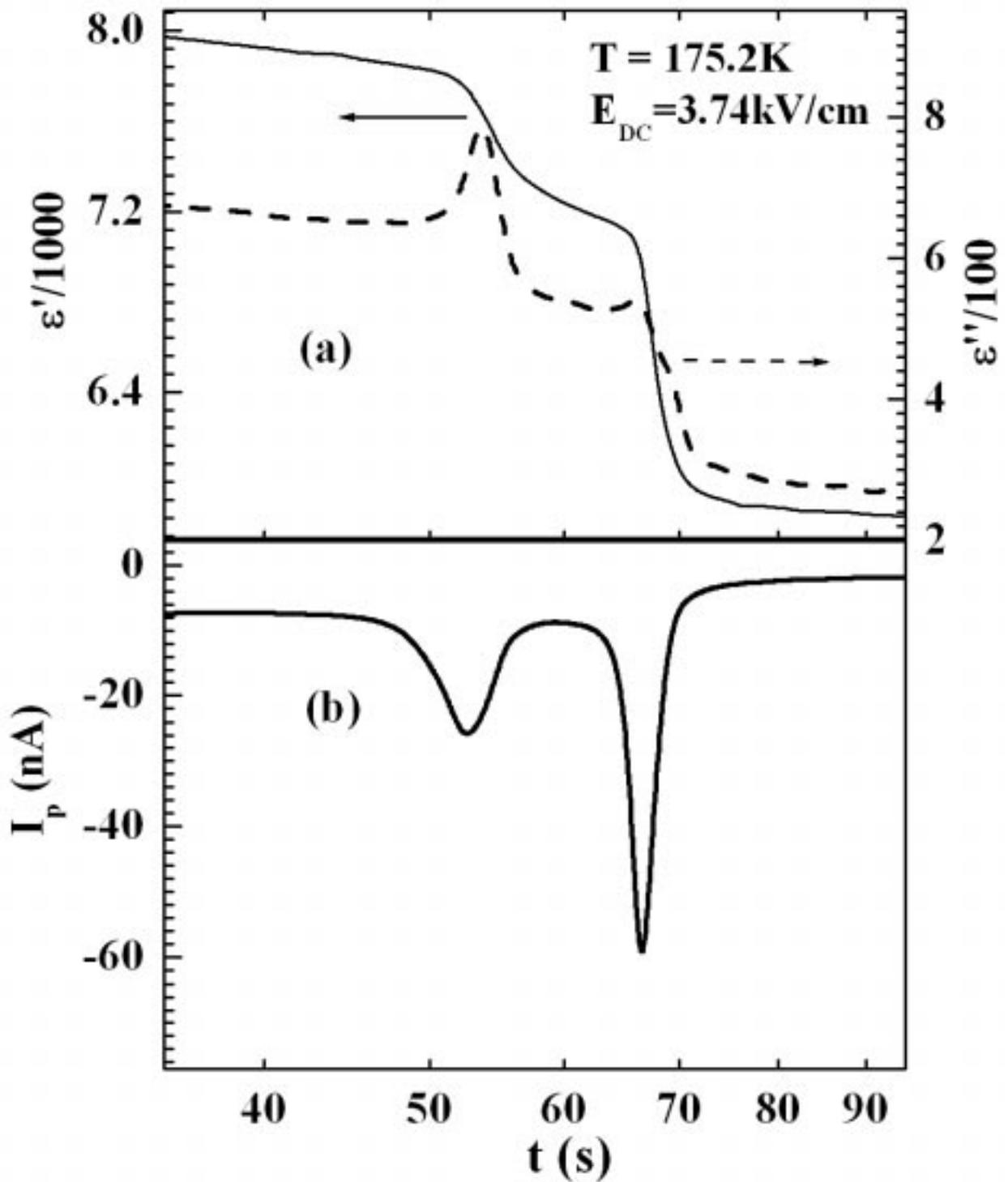

Figure 2. ε'(t), ε"(t), and $I_P(t)$ after applying a field of 3.7 kV/cm (at t=0)at 175 K after zero-field cooling.

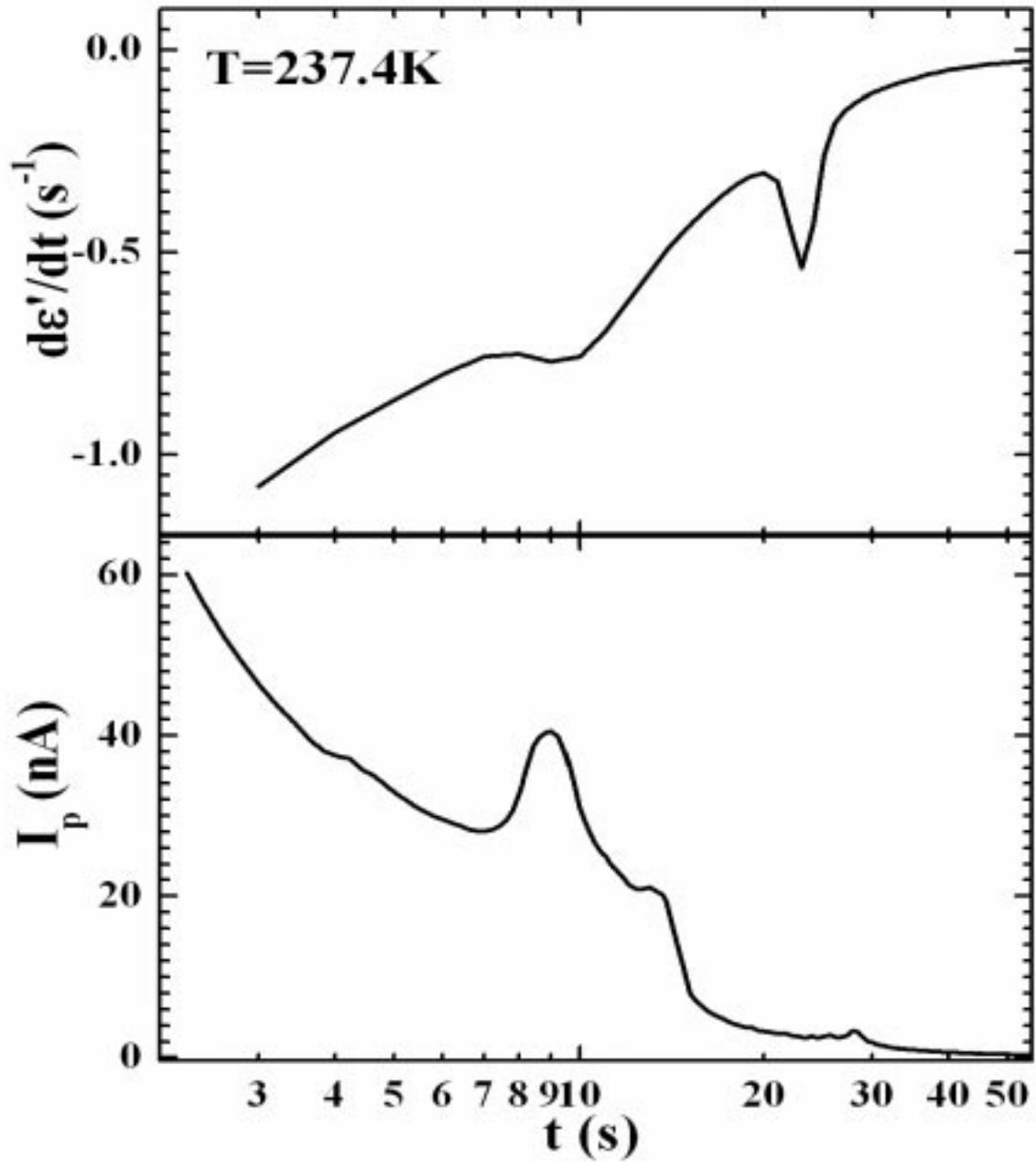

Figure 3. dε'(t)/dt and the polarization current are shown as a function of time t after switching on a field of 2.5 kV/cm for the [001] PMN-6%PT sample.

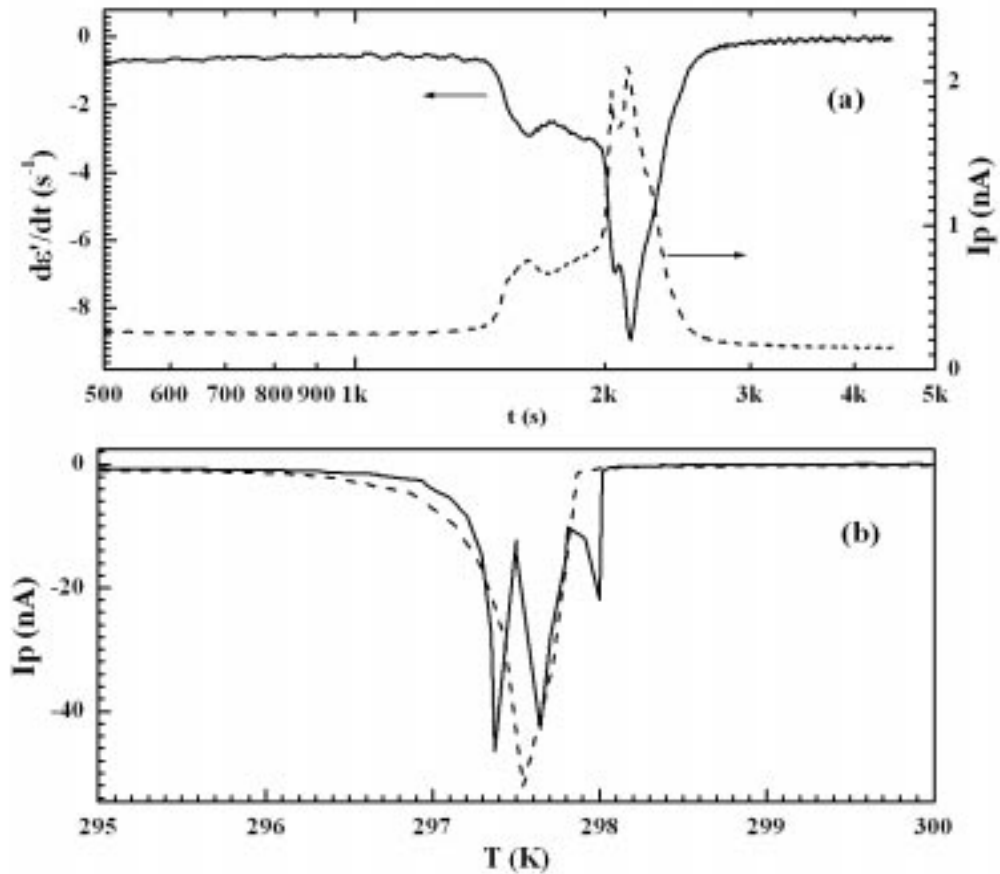

Figure 4. Part (a) shows $d\varepsilon'(t)/dt$ (solid) and $I_p(t)$ (dashed) after switching on E=0.55 kV/cm at T= 250 K for the [111] PMN-12%PT sample. Part (b) shows $I_p(T)$ (solid) on warming at E=0 the ZFC-F2 state, with the corresponding data (dashed) for a simple FC prepared state. (The FC data were taken at half the sweep rate, and then rescaled to approximately compensate.)

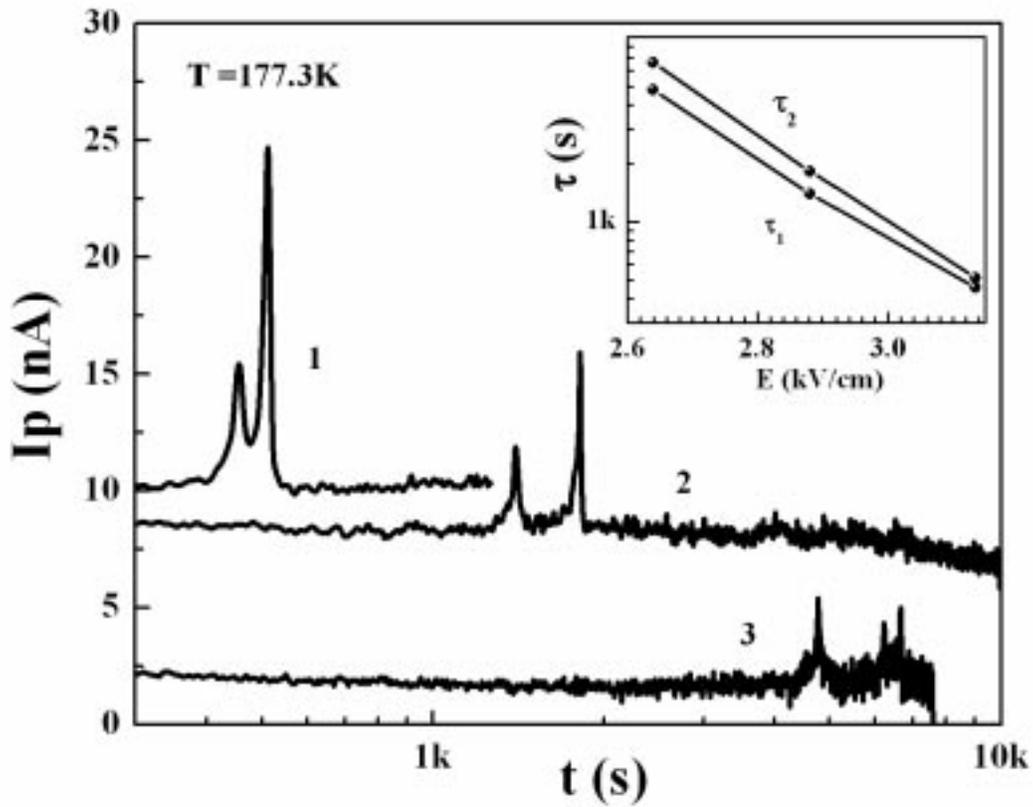

Figure 5. The two-step pulses in $I_p(t)$ are shown for different E at 177K. The curves have been offset for clarity. Curve 1) was taken at 3.17 kV/cm, curve 2) at 2.91 kV/cm (with $I_P$ scaled x2), and curve 3) at 2.67 kV/cm (with $I_P$ scaled x 5)The inset shows the E-dependences of the times.

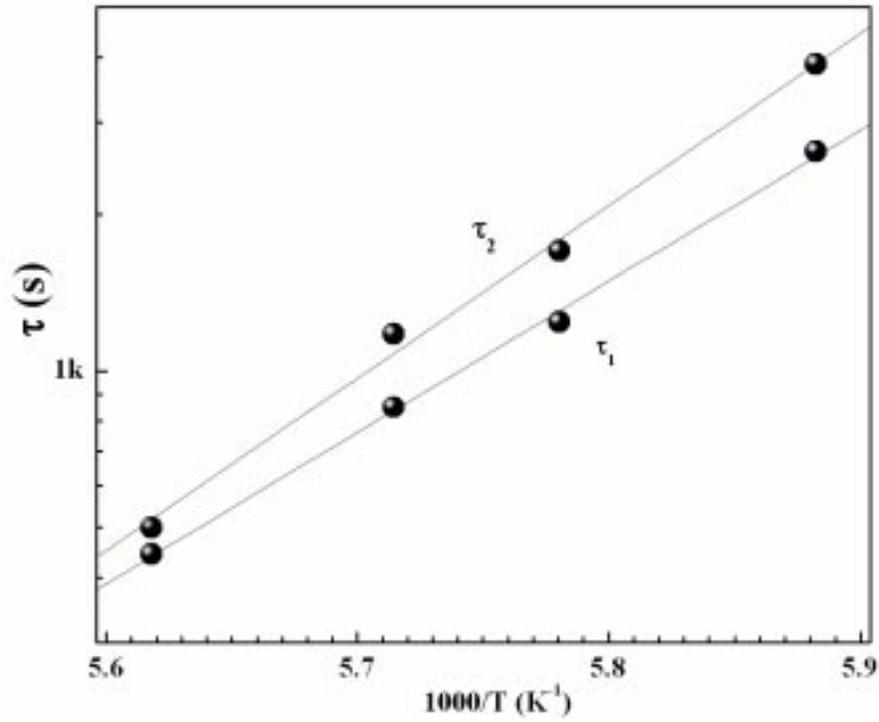

Figure 6. Arrhenius plots of the T-dependences of the two times are shown at E= 3.17 kV/cm. The fits give activation energies of 6700 K and 7600 K with nominal attempt rates of $5 \times 10^{13}$ s$^{-1}$ and $8 \times 10^{15}$ s$^{-1}$ respectively.

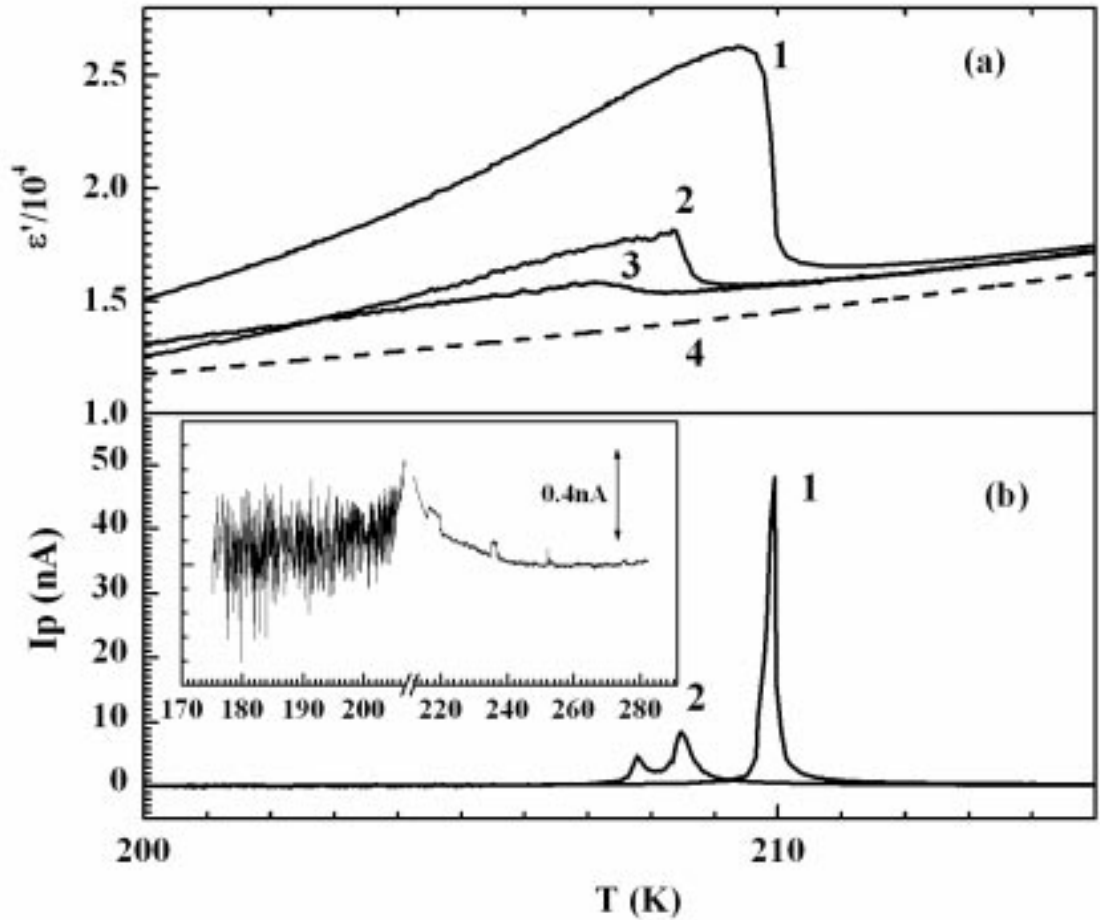

Figure 7. (a) ε'(T) is shown on zero-field warming after different E-T histories: (1) FC at 3.74 kV/cm. (2) ZFC-F2 prepared at 3.74 kV/cm and 175 K, (3) as in (2) but ZFC-F1, interrupted after the first step, and (4) ZF, simple zero-field cycling. (b) $I_P(T)$ is shown for procedures (1) and (2). The inset shows a blow-up of curve (2), in which the large noise is evident. Curve (1) is not perceptibly noisy on this scale.

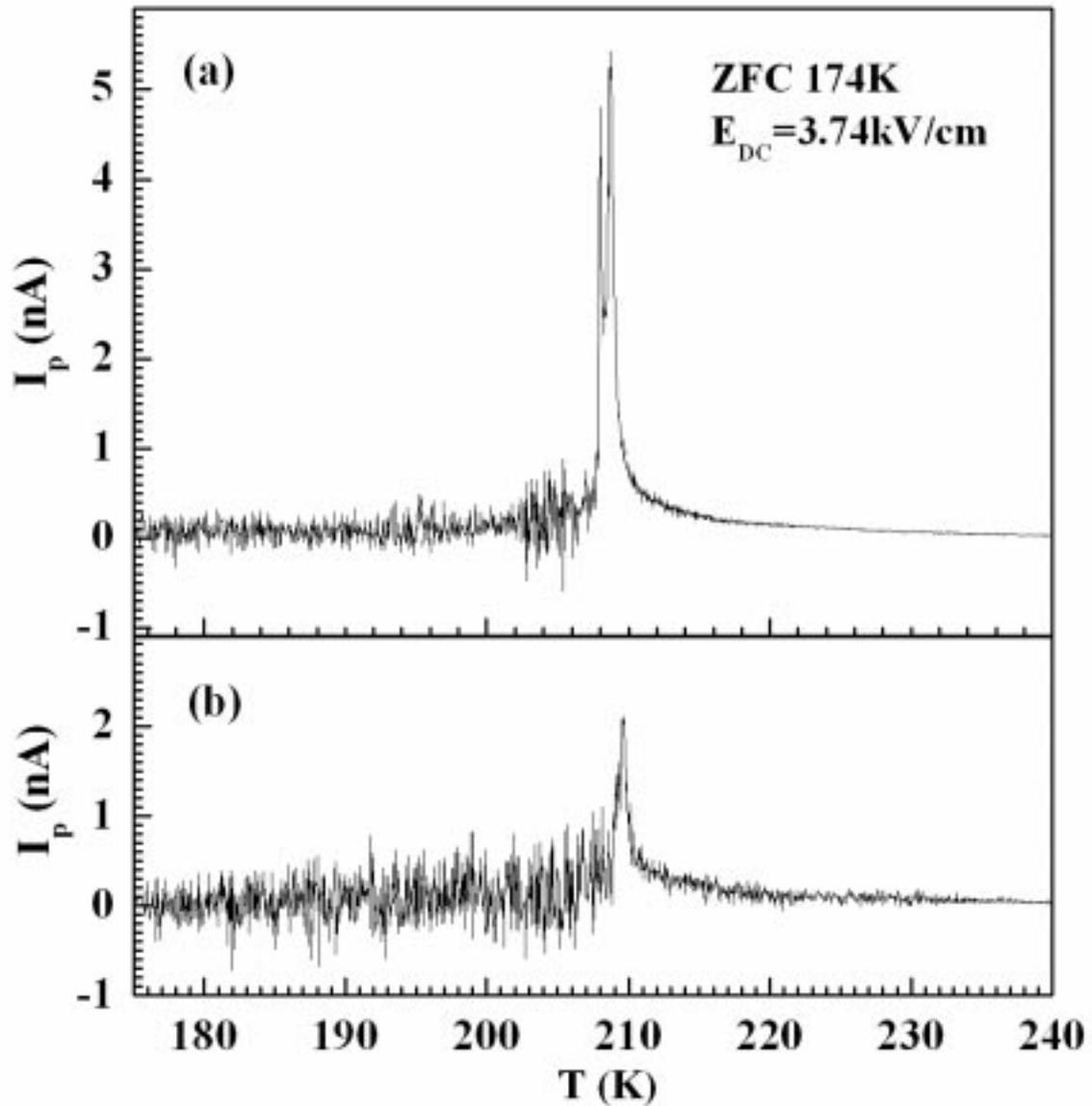

Figure 8. $I_p(T)$ on zero-field warming is shown in detail for (a) ZFC-F2 (b) ZFC-F1, illustrating both that the approximate reproducibility of the large current noise of ZFC-F2 in the previous figure, and the larger noise from ZFC-F1. This ZFC-F1 warming curve was taken after 17 hrs of aging at E=0 and T= 174 K, illustrating the stability of this state. Another curve taken after very brief aging looks very similar.